\documentclass{aa}  

\usepackage{graphicx}
\usepackage{txfonts}
\usepackage{lipsum}
\usepackage{textcomp}
\usepackage{subcaption}
\usepackage{lscape}             
\usepackage{placeins}

\begin{document}

   \title{Revisiting candidate high-velocity stars associated with the Sagittarius dwarf spheroidal galaxy}

   \author{Jian Zhang\inst{1} \and
        Cuihua Du\inst{1} \and
        Mingji Deng\inst{1} \and
        Haoyang Liu\inst{1} \and
        Zhongcheng Li\inst{1}
        }

\institute{School of Astronomy and Space Sciences,
           University of Chinese Academy of Sciences,
           Beijing 100049, People’s Republic of China
           \email{ducuihua@ucas.ac.cn}
           }
 
  \abstract
   {Hypervelocity stars (HVSs) are valuable tracers of extreme dynamical processes. The Sagittarius dwarf spheroidal galaxy (Sgr dSph), currently undergoing tidal disruption, offers a unique environment to search for such stars.}
   {We aim to identify candidate HVSs dynamically linked to the Sgr dSph and to assess their possible origins.} 
   {Using Gaia DR3, DESI DR1, and LAMOST DR12, we selected stars with galactocentric velocities above 400 km\,s$^{-1}$ and traced their orbits in a realistic Galactic potential including the Sgr dSph and the Large Magellanic Cloud. We then tested three scenarios for their origin: the Hills mechanism, tidal disruption, and random halo star encounters.}
   {We identified 95 candidates passing within 2.5 half-mass radii of the Sgr dSph. Their kinematics are inconsistent with production by the Hills mechanism or tidal disruption but are well reproduced by halo stars that naturally cross the Sgr orbit. Furthermore, their metallicity distribution is consistent with that of the Milky Way halo rather than the Sgr stream or Sgr dSph.}
   {Our results suggest that our candidates and those in previous studies are most likely halo stars rather than genuine Sgr-origin HVSs. This highlights the need to account for the halo population when inferring stellar origins from orbital analysis and that chemical abundances will be a valuable constraint in the future. While we detect no unbound Sgr HVSs, such a discovery would directly imply extreme dynamical processes. Our results serve as a basis for future studies with upcoming surveys.}


   \maketitle

\nolinenumbers

\section{Introduction}
In the Galactic stellar halo, most stars exhibit random motions with velocities of $100-150~\mathrm{km~s^{-1}}$. However, some high-velocity stars travel significantly faster—on the order of several hundred kilometers per second—with a subset known as hypervelocity stars (HVSs) even exceeding the local escape speed. Such objects are indicative of extreme dynamical processes. The classic Hills mechanism \citep{Hills1988} was the first to predict the existence of HVSs. In this scenario, when a stellar binary passes sufficiently close to the massive black hole (MBH) at the Galactic center (GC), tidal forces disrupt the binary. Ultimately, one component is captured by the MBH, while its companion is ejected at speeds potentially exceeding $1000~\mathrm{km~s^{-1}}$.

Subsequent work has shown that single stars can also achieve extremely high velocities through interactions with more complex MBH systems. For example, encounters between a star and a binary MBH or an MBH–intermediate-mass black hole (IMBH) pair can efficiently transfer energy to the star, ejecting it at hundreds to thousands of kilometers per second \citep{Yu2003,Gualandris2009,Rasskazov2019}. Similarly, a dense cluster of stellar-mass black holes orbiting the central MBH can scatter passing stars so that they become HVSs \citep{O'Leary2008}. Infalling globular clusters, when tidally disrupted by the GC’s MBH (or an MBH binary), can also eject stars at extreme velocities \citep{Capuzzo-Dolcetta2015,Fragione2017}.

Supernovae in binary systems constitute another natural channel for producing fast runaway stars. Core-collapse supernovae disrupt their progenitor binaries and can impart kicks of up to $300-400~\mathrm{km~s^{-1}}$ \citep{Evans2020}, accounting for many OB runaway stars found above the Galactic plane \citep{Blaauw1961,Portegies2000}. Thermonuclear supernovae, in contrast, can accelerate companions to much higher speeds. In particular, detonations in white dwarf–white dwarf binaries or in systems pairing a white dwarf with a nondegenerate star have been proposed to yield velocities exceeding $1000~\mathrm{km~s^{-1}}$ \citep{Shen2018,Wang2009,Bauer2019,Neunteufel2020, El-Badry2023}.

Dense stellar environments and galactic interactions provide alternative channels for hypervelocity ejections. Close multi-body encounters within young massive clusters can scatter stars onto unbound trajectories \citep{Perets2012, Oh2016}. Specifically, three-body interactions---particularly single star-binary encounters involving compact objects---can eject stars at velocities reaching several hundred kilometers per second \citep{Cabrera2023, Weatherford2023, Grondin2023, Grondin2024, Evans2025}. Additionally, the tidal disruption of infalling dwarf galaxies can strip and accelerate stars to extreme velocities \citep{Abadi2009, Piffl2011}. Collectively, these mechanisms illustrate the diverse dynamical processes capable of populating the Galactic halo with HVSs.

The first candidate HVS was identified by \citet{Brown2005}. It was a B-type star traveling at a total velocity of $709~\mathrm{km~s^{-1}}$ through the Galactic halo. After that, over two dozen HVSs, all early-type, have been identified either by incidental discoveries \citep{Hirsch2005, Heber2008, Koposov2020} or follow-up surveys \citep{Brown2014}. More recently, the Southern Stellar Stream Spectroscopic Survey (S5) reported the fastest known HVS to date \citep{Koposov2020}. This A-type star moves at $1755~\mathrm{km~s^{-1}}$ and lies at a heliocentric distance of approximately $9~\mathrm{kpc}$, with a GC-related origin according to orbital tracing.

The advent of the \textit{Gaia} mission has revolutionized the search for HVSs by providing accurate astrometry. Thanks to \textit{Gaia} Data Releases 2 and 3 \citep{Gaia2018,Gaia2021}, dozens of new HVS candidates have been uncovered (e.g., \citealt{Bromley2018}; \citealt{Du2018,Du2019}; \citealt{Shen2018}; \citealt{Marchetti2019,Marchetti2021,Marchetti2022}; \citealt{Liao2023}). In addition to \textit{Gaia}, large spectroscopic surveys have also proven to be powerful tools for discovering HVSs. For instance, \citet{Sun2025} recently identified ten new unbound B- and A-type HVS candidates using the Large Sky Area Multi-Object Fiber Spectroscopic Telescope (LAMOST).

The Sagittarius dwarf spheroidal galaxy (Sgr dSph) is the closest and third most massive satellite of the Milky Way, and it is currently undergoing significant tidal disruption \citep{Ibata94}. The HVS and high-velocity star candidates originating from the Sgr dSph provide a unique window into its central dynamics and tidal history. \citet{Huang2021} reported the first such candidate, J1443+1453, a star with a galactocentric velocity of $599~\mathrm{km~s^{-1}}$ that likely originated from the Sgr dSph. 
Building on this, \citet{LiH2022} (hereafter Li22) combined proper motions from Gaia Early Data Release 3 (Gaia EDR3; \cite{Gaia2021}) with radial velocities from multiple spectroscopic surveys to identify 60 high-velocity candidates whose orbits are consistent with a Sgr origin. The locations of these stars on the Hertzsprung–Russell diagram, coupled with the chemical properties of the 19 candidates with measured metallicities, show similarities to members of the Sgr stream. The closest approach of the Li22 candidates to the Sgr dSph occurred during the progenitor's pericentric passage ($\sim$38.2 Myr ago), leading the authors to suggest that these stars were tidally stripped from the system. Additionally, Li22 proposed that two HVSs within their sample might have originated from the Hills mechanism. 
Furthermore, \citet{LiQ2023} (hereafter Li23) assembled a large sample of 547 extreme-velocity stars with galactocentric velocities $V_{\mathrm{GSR}} > 0.8 V_{\mathrm{esc}}$. Among these, 15 stars---including two HVS candidates---feature orbits that trace back to close encounters with the Sgr dSph. This hypothesized Sgr origin is further supported by an analysis of their distribution in the [$\alpha$/Fe]--[Fe/H] plane.

Although many observational discoveries have been made, the dynamical mechanisms responsible for ejecting these stars have not been thoroughly explored. In this work, we investigate three possible origins for high-velocity stars linked to the Sgr dSph: the classic Hills mechanism, tidal stripping during the dwarf's ongoing disruption, and chance passages of high-velocity halo stars through the Sgr dSph.
The paper is structured as follows. In Section \ref{sec:sample}, we describe our selection of high–velocity stars from Gaia DR3, DESI DR1, and LAMOST DR12. In Section \ref{sec:origin}, we examine three possible origins for these stars:  ejection via the Hills mechanism (Section \ref{subsec:Hills}), tidal stripping during the dwarf's disruption (Section \ref{subsec:Tidal}), and chance encounter of high-velocity halo stars with the Sgr dSph (Section \ref{subsec:halo}). We provide a complementary chemical abundance analysis in Section \ref{sec:Chemical}. Finally, we present a concise discussion and our conclusions in Section \ref{sec:D&C}.

\section{Sample selection} \label{sec:sample}
\subsection{Data} \label{subsec:data}
The Gaia DR3 catalog provides astrophysical parameters for approximately 470 million stars \citep{Gaia2023} and radial velocities for 34 million sources in the Gaia DR3 Radial Velocity Spectrometer (RVS) measured sample \citep{Katz2023}. To ensure high‐quality astrometric measurements, we applied the following criteria: 

(1) Renormalized unit weight error $\mathtt{RUWE} < 1.4$. Small values mean the source is well fitted by a single star model. 

(2) $\mathtt{ASTROMETRIC\_EXCESS\_NOISE\_SIG} < 2$. Excess noise refers to the extra noise in each observation that causes the residual scatter in the astrometric solution. If $\mathtt{ASTROMETRIC\_EXCESS\_NOISE\_SIG}$ is greater than two, then this excess noise cannot be ignored.

(3) $\mathtt{ASTROMETRIC\_GOF\_AL} < 3$. This parameter represents the goodness of fit between the astrometric model and the observed data. A higher value indicates a poor fit.

(4) $\mathtt{VISIBILITY\_PERIODS\_USED} > 10$. This parameter represents the number of visibility periods used in the astrometric solution. A high number of visibility periods is a better indicator of an astrometrically well–observed source.

(5) $\varpi - \varpi_{ZP} > 5\,\sigma_{\varpi}$. Here $\varpi$ is the parallax and the $\sigma$ terms denote its uncertainty. Parallax in the Gaia DR3 are known to exhibit a small systematic offset. We therefore corrected the reported parallaxes $\varpi$ by subtracting the zero-point offset $\varpi_{ZP}$ following the procedure of \citet{Lindegren2021a}.

(6) For the Gaia DR3 RVS sample, we further included the following two criteria to ensure a robust radial velocity measurement: $\mathtt{RV\_EXPECTED\_SIG\_TO\_NOISE} \ge 5$ and $\mathtt{RV\_NB\_TRANSITS} \ge 10$. These cuts remove sources with low signal-to-noise radial velocity measurements and sources with an insufficient number of RVS transits.

Next, coordinates of selected Gaia DR3 RVS sample stars were transformed to the Galactocentric Cartesian frame with their 6D phase information by \textit{Astropy} \citep{Astropy2022}. The heliocentric distance $d_\odot$ is equal to $1 / (\omega - \omega_{ZP})$, the Sun's position $(x_\odot, y_\odot, z_\odot) = (8.122,\;0,\;0.0208)\,\mathrm{kpc}$ \citep{GRAVITY2018, Bennett2019} was set, and the velocity $(U_\odot, V_\odot, W_\odot) = (11.1,\;245,\;7.25)\,\mathrm{km\,s^{-1}}$ \citep{Sch"onrich2010, McMillan2017} was adopted. Then we imposed a velocity selection criterion of total velocity $v_{\rm GC} > 300\ \mathrm{km\,s^{-1}}$. By combining these velocity thresholds with the previous quality cuts, we expected to retain only genuine high-velocity stars.

The DESI DR1 provides spectra for over 18 million unique targets observed in the Main Survey between May 2021 and June 2022 \citep{DESI2025}. To enhance the scientific usability beyond the core spectroscopic outputs, the DESI Science Collaboration has produced a suite of value-added catalogs (VACs) accompanying DR1 \citep{Koposov2025}.
In this study, we employed the Milky Way Survey (MWS) catalog, which contains detailed analyses of stellar spectra produced by the MWS Working Group, including stellar parameter determinations and quality assessments \citep{Koposov2025}. We combined radial velocities and quality flags from the MWS catalog with astrometric parameters and their uncertainties from Gaia DR3 through cross-matching.

To ensure a robust kinematic sample, we required \texttt{RVS\_WARN} = 0, which indicates that the stars have reliable stellar model fits by the \texttt{RVS} pipeline.
Finally, following the procedure applied to the Gaia DR3 RVS sample, we transformed the DESI DR1 MWS stars into Galactocentric coordinates and imposed a selection criterion of total velocity $v_{\rm GC} > 300\ \mathrm{km\,s^{-1}}$.

LAMOST is a four-meter-aperture reflecting Schmidt telescope designed for efficiently obtaining spectra of both Galactic and extra-galactic objects in the optical band \citep{Cui2012, Zhao2012}. We utilized the A-, F-, G-, and K-type stellar parameter catalogs from the LAMOST DR12 low-resolution survey (LRS), which comprise a total of 8,370,041 spectra. Of these, 471,029 are A-type spectra, while the F-, G-, and K-type stars comprise 2,412,434; 3,966,183; and 1,520,395 spectra, respectively. Specifically, we used the radial velocities in LAMOST LRS DR12 derived by the LAMOST Stellar Parameter Pipeline (LASP; \cite{Luo2015}), and we crossmatched other astrometric parameters with Gaia DR3.
We applied the same quality filters as for the Gaia DR3 sample to the crossmatched LAMOST DR12–Gaia DR3 data and retained only sources with galactocentric total velocities $v_{\mathrm{GC}} > 300~\mathrm{km~s^{-1}}$.

\subsection{Orbital analysis} \label{subsec:orbits}

To properly account for measurement uncertainties, we adopted a Bayesian framework. The posterior probability of the parameters
$\theta = (d,\;v_\alpha,\;v_\delta,\;v_r)$ given the observables
$x = (\omega,\;\mu_\alpha^*,\;\mu_\delta,\;rv)^{T}$
is proportional to the product of the likelihood, which is modeled as a three-dimensional Gaussian distribution, and the prior probability:
\begin{equation}\label{eq:posterior}
  P(\theta\mid x)\;\propto\;\exp\!\biggl[-\tfrac{1}{2}\bigl(x - m(\theta)\bigr)^{T}C_{x}^{-1}\bigl(x - m(\theta)\bigr)\biggr]\;\times\;P_{\rm prior}(d|\alpha, \beta, L),  
\end{equation}
where
$$
m(\theta)=\bigl(1/d,\;v_\alpha/(k\,d),\;v_\delta/(k\,d),\;v_r\bigr)^{T},\quad
k = 4.74047,
$$
and $C_x$ is the covariance matrix of the observables.
For the distance prior $P_{\rm prior}(d|\alpha, \beta, L)$, we used the geometric distance prior of \citet{Bailer-Jones2021}, which is a three-parameter generalized gamma distribution and parameters are determined by each star's HEALPix index.

We implemented Markov chain Monte Carlo sampling using the Python package \textit{emcee} \citep{Foreman13} to draw samples from the posterior distribution, generating 5,000 Monte Carlo (MC) realizations for each star. Next, each realization’s coordinates and velocities were transformed to a Galactocentric frame. For each star, we took the median (50th percentile) of these realizations as the central estimate, with uncertainties defined by the 16th and 84th percentiles. In this work, we define high-velocity stars as those with galactocentric velocities greater than 400 $\mathrm{km~s^{-1}}$, and HVSs as those with galactocentric velocities exceeding 600 $\mathrm{km~s^{-1}}$. In previous studies, HVSs were commonly defined based on their escape probability \citep{Marchetti2019, Marchetti2021, Liao2023}, determined by comparing a star’s velocity with the local escape velocity; stars with escape probabilities greater than 50\% were classified as HVSs. We argue that this definition makes the classification strongly dependent on distance uncertainties, since stars with larger distance errors are more likely to be assigned high escape probabilities. Instead, we adopted a purely velocity-based definition, which we consider more robust. Subsequently, we compare our HVS candidates with the escape velocity curve to assess whether they are truly unbound.

For the high-velocity star candidates, we integrated their MC orbital realizations backward in time within a time-dependent Galactic potential. This potential consists of the static \texttt{MWPotential2014} \citep{Bovy2015} together with the moving Large Magellanic Cloud (LMC) and the moving Sgr dSph.

The LMC was modeled by a Hernquist potential \citep{Hernquist1990} with a total mass of $M_{\mathrm{LMC}} = 1.5\times10^{11}~M_\odot$ and a scale radius of $r_{\mathrm{LMC}} = 17.14~\mathrm{kpc}$ \citep{Erkal2019, Erkal2020}. At $t=0$ (present day) its center of mass is placed at $(\alpha,\delta) = (78.76^{\circ}, -69.19^{\circ})$ with a heliocentric distance of $49.59\mathrm{kpc}$ \citep{Zivick2019, Pietrzynski2013}, a proper motion of $(\mu_\alpha^\ast,\mu_\delta) = (1.91, 0.229)~\mathrm{mas,yr^{-1}}$, and a heliocentric line-of-sight velocity of $262.9~\mathrm{km~s^{-1}}$ \citep{Van2002, Kallivayalil2013}. We integrated the LMC orbit 1 Gyr into the past in \texttt{MWPotential2014}, including dynamical friction following \citet{Chandrasekhar1943}, and represented its time-varying contribution with the \texttt{MovingObjectPotential} built in \textit{galpy}.

We modeled the Sgr dSph using a Plummer potential. To satisfy the observational constraints from \citet{Vasiliev2020}, who reported a mass of $4\times10^{8}~M_\odot$ within 5 kpc and a peak circular velocity of $21~\mathrm{km~s^{-1}}$ at $\sim$2.5--3~kpc, we adopted the structural parameters from \cite{Rostami2024}: $M_{\mathrm{Sgr}} = 4.8\times10^{8}~M_\odot$, a scale radius of $r_{\mathrm{sc}} = 1.8~\mathrm{kpc}$, and a half-mass radius of $R_{\mathrm{h}} = 2.34~\mathrm{kpc}$. The present-day Sgr dSph remnant is positioned at $(\alpha, \delta) = (283.76^{\circ}, -30.48^{\circ})$, with proper motions of $(\mu_\alpha^\ast, \mu_\delta) = (-2.70, -1.35)~\mathrm{mas~yr^{-1}}$ and a heliocentric line-of-sight velocity of $142~\mathrm{km~s^{-1}}$ \citep{Vasiliev2020, Vasiliev2021}. The distance to the Sgr center remains the primary source of kinematic uncertainty, with values in the literature ranging from 24 to 28~kpc \citep{Ibata1997, Hamanowicz2016, Ferguson2020, Vasiliev2020, Vasiliev2021}. Our tests indicate that while a smaller distance increases the number of high-velocity stars undergoing close encounters (as detailed in Section \ref{subsec:halo}), the choice of distance does not significantly skew the resulting kinematic distributions (e.g., encounter velocities or flight times). Consequently, we adopted a distance of 27~kpc \citep{Vasiliev2021}, as it provides an optimal fit to the observed stream properties. We integrated the Sgr dSph orbit backward in time for 1~Gyr, accounting for dynamical friction and the gravitational influence of both the Milky Way and the LMC.

\begin{figure}[htbp]
    \centering
    \includegraphics[width=\linewidth]{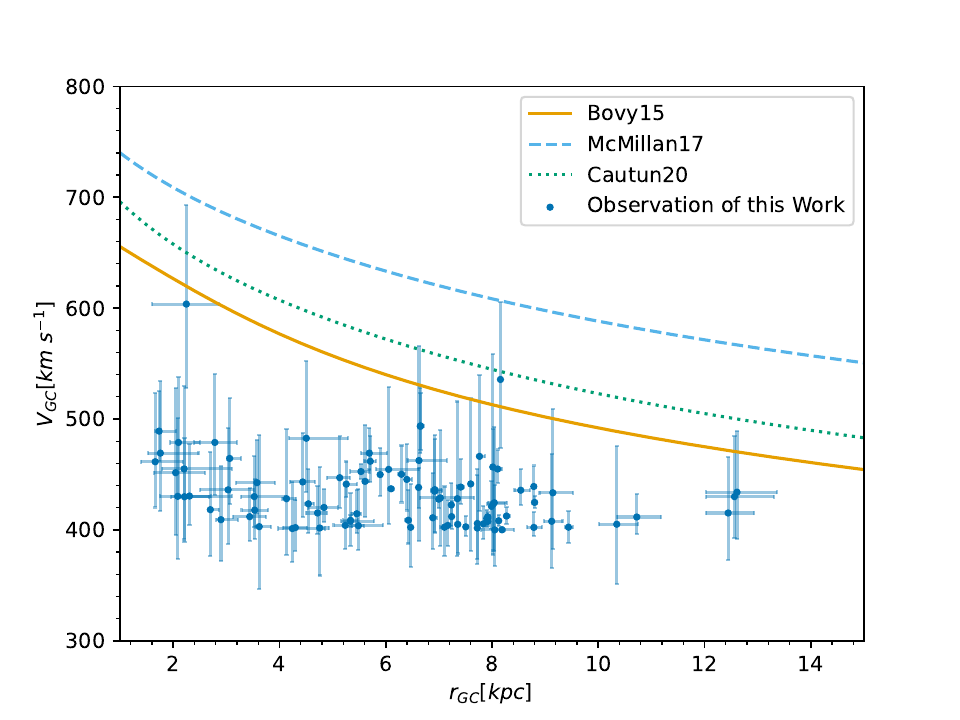}
    \caption{Distribution of galactocentric total velocity ($v_{\mathrm{GC}}$) as a function of galactocentric radius ($r_{\mathrm{GC}}$) for high-velocity stars ($P_{\mathrm{Sgr}}>0.5$). Solid, dashed, and dotted curves show the escape-velocity profiles predicted by three Milky Way potential models: \cite{Bovy2015}, \cite{McMillan2017}, and \cite{Cautun2020}.
}
    \label{fig:vgc_over_rgc}
\end{figure}

We quantified the probability that a given star experiences a close encounter with the Sgr dSph, denoted as $P_{\mathrm{Sgr}}$, by
\begin{equation}
       P_{\mathrm{Sgr}}=\frac{N_{\mathrm{MC}}\bigl(r_{\min}<2.5\,R_h\bigr)}{N_{\mathrm{MC}}},
\end{equation}
where $r_{\min}$ is the closest distance between the star and the Sgr dSph along its backward-integrated orbit, $N_{\rm MC} = 5000$ is the number of MC realizations used in this study, and $N_{\rm MC}(\mathtt{condition})$ is the number of MC realizations that satisfy the corresponding condition. The threshold $2.5R_h$ corresponds to two and a half times the remnant’s half-mass radius, providing a physically motivated boundary.
In Gaia DR3, we identify 60 stars with $P_{\mathrm{Sgr}}>0.5$ out of 2967 high-velocity stars (2.02\%), including six sources also reported by Li22. DESI DR1 contributes 17 such stars among 732 (2.32\%), of which one overlaps with Li23. LAMOST DR12 contributes 18 such stars among 524 (3.44\%), including five in common with Li23. These stars are displayed in Figure \ref{fig:vgc_over_rgc}. We find only one star with $v_\mathrm{GC}$ exceeding $600~\mathrm{km~s^{-1}}$; however, its galactocentric distance is about 2.25 kpc, and this velocity lies below the local escape velocity curve. We therefore conclude that no HVS originating from the Sgr dSph are identified in this work.

\begin{figure}[htbp]
    \centering
    \includegraphics[width=\linewidth]{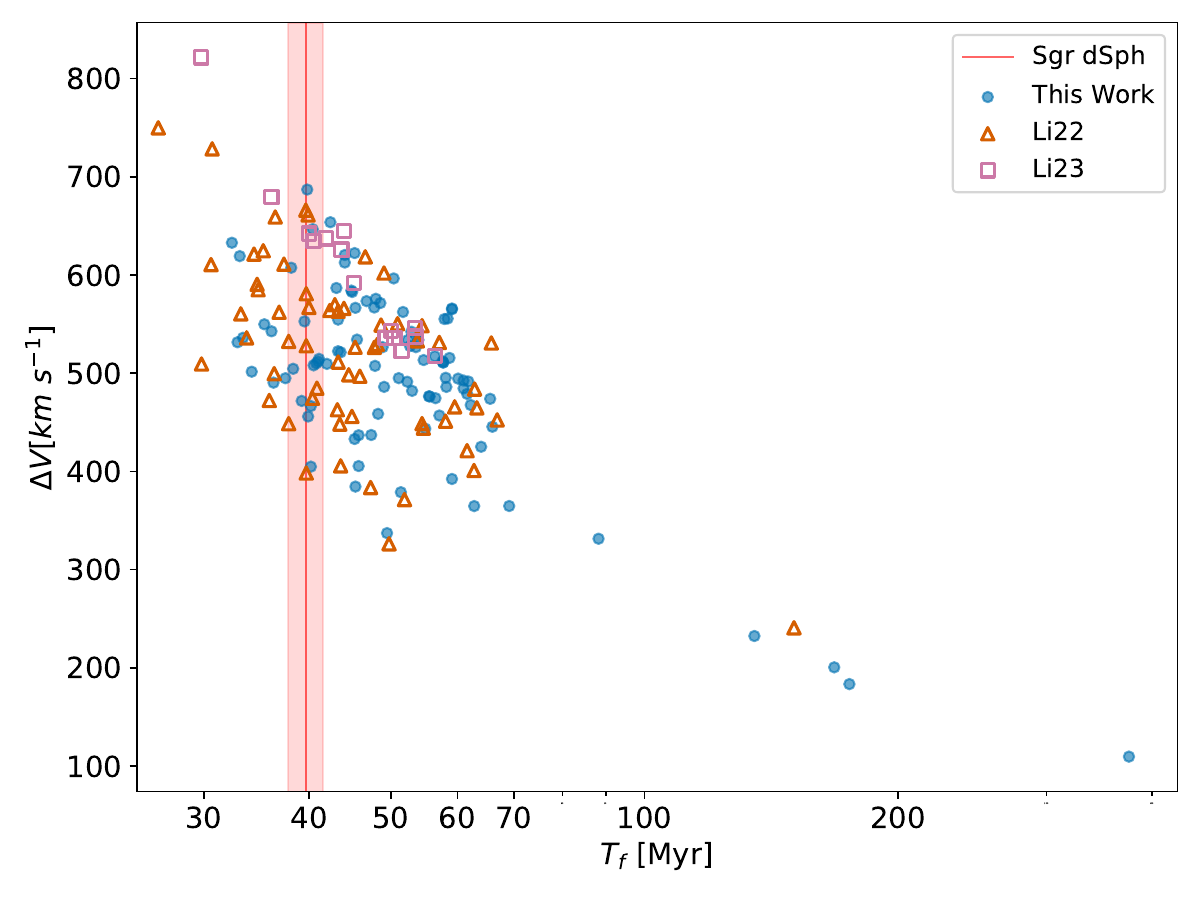}
    \caption{Relative velocity $\Delta V$ with respect to the Sgr dSph at pericentric passage versus flight time $T_f$ on a logarithmic scale. Blue circles, orange triangles, and magenta squares represent results from this work, Li22, and Li23, respectively. Vertical red line indicates the 39.7 Myr flight time of the Sgr dSph since its most recent pericentric passage, with the shaded region representing the $1\sigma$ confidence interval.}
    \label{fig:tf_and_dv}
\end{figure}

For each high-velocity star with $P_{\mathrm{Sgr}}>0.5$, we computed the relative velocity with respect to the Sgr dSph at pericentric passage ($\Delta V$) and the corresponding flight time to the present ($T_f$). The results are illustrated in Figure \ref{fig:tf_and_dv}.

A clear clustering is observed for all three samples around $T_f \sim 30$--$70~\mathrm{Myr}$ and $\Delta V \sim 400$--$600~\mathrm{km~s^{-1}}$, consistent with the flight time of the Sgr dSph ($\sim 39.7~\mathrm{Myr}$) since its most recent pericentric passage \citep{LiQ2023}. However, distinct differences appear in the extended distributions. Our sample exhibits a significant population extending towards long flight times ($T_f \gtrsim 100~\mathrm{Myr}$) and correspondingly low relative velocities ($\Delta V < 300~\mathrm{km~s^{-1}}$). We interpret these outliers as halo stars, which is discussed further in Section \ref{subsec:halo}. In comparison, Li22 shows a similar but sparser trend in this high-$T_f$/low-$\Delta V$ regime, while Li23 is tightly confined to the region associated with the recent Sgr passage.

\begin{figure*}[htbp]
    \centering
    \includegraphics[width=\linewidth]{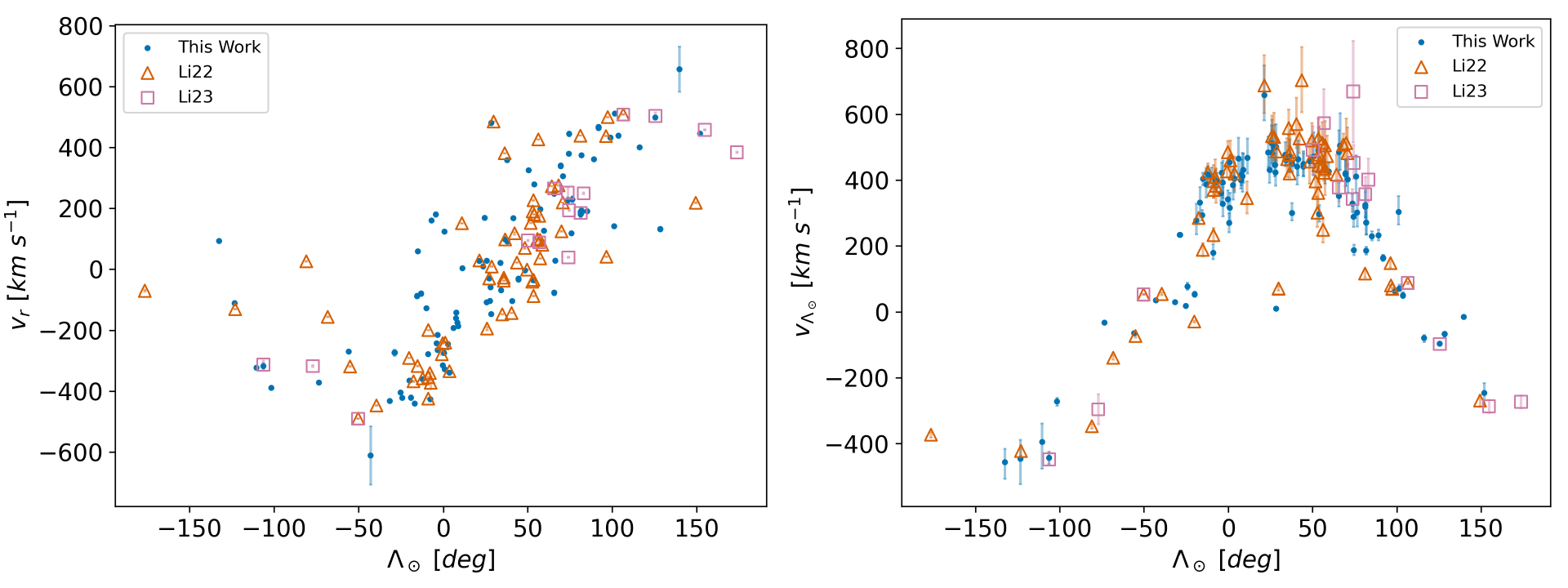}
    \caption{Distribution of high-velocity stars in Sgr Stream coordinates \citep{Vasiliev2021}. Left panel: Radial velocity over Sgr stream coordinates $\Lambda_\odot$. Right panel: Tangential velocity in the $\Lambda_\odot$ direction over $\Lambda_\odot$. Blue circles, orange triangles, and magenta squares represent results from this work, Li22, and Li23, respectively.
}
    \label{fig:rv_and_vL_over_L}
\end{figure*}

Since the high-velocity stars in our sample experienced recent close encounters with the Sgr dSph, their motions are expected to be intrinsically linked to the dynamics of the Sgr system. To better characterize these kinematic patterns and facilitate a direct comparison with the Sgr stellar stream, we transformed the heliocentric coordinates of our candidates into the Sgr stream coordinate system $(\Lambda_{\odot}, B_{\odot})$. Given the variety of conflicting conventions in the literature, we adopted the right-handed heliocentric system as defined in \cite{Vasiliev2021} to ensure consistency with standard Galactic and International Celestial Reference System frames. Specifically, we followed the convention of \cite{Belokurov2014} for the longitudinal coordinate $\Lambda_{\odot}$, which represents the angle along the Sgr orbital plane. In this system, $\Lambda_{\odot}$ increases toward the leading arm ($\Lambda_{\odot} > 0$) and decreases toward the trailing arm ($\Lambda_{\odot} < 0$). For the latitude $B_{\odot}$, defined perpendicular to the orbital plane, we followed the convention of \cite{Majewski2003}, which ensures that the orbital pole of the Sgr plane lies at $B_{\odot} = 90^\circ$ (corresponding to $(l, b) = (273.75^\circ, -13.46^\circ)$ in Galactic coordinates, or $(\alpha, \delta) = (123.65^\circ, -53.52^\circ)$ in the International Celestial Reference System). In this coordinate system, the Sgr dSph remnant is centered at $(\Lambda_{\odot, 0}, B_{\odot, 0}) \approx (0^\circ, 1.5^\circ)$. We plot those high-velocity stars in Figure \ref{fig:rv_and_vL_over_L}. In the left panel, the line‐of‐sight velocities $v_r$ of these high-velocity stars display a clear correlation with the stream longitude $\Lambda_\odot$. The distribution is symmetric about $\Lambda_\odot = 50^\circ$: for $\Lambda_\odot > 50^\circ$, $v_{\rm r} > 0$ (receding from the Sun), whereas for $\Lambda_\odot < 50^\circ$, $v_{\rm r} < 0$ (approaching the Sun). This behavior arises from projection effects along the line of sight, since the stars share a similar orbital phase near pericentric passage with Sgr dSph—some having passed the Solar azimuth produce positive $v_r$, while others moving toward the Sun produce negative values. 

In the right panel, the tangential velocity component $v_{\Lambda_\odot}$ measured along the stream longitude exhibits a cosine‐like variation with $\Lambda_\odot$, reaching its maximum at $\Lambda_\odot = 50^\circ$ where the line‐of‐sight component crosses zero. This sinusoidal pattern likewise reflects geometrical projection of the common star motion onto the tangential direction. All three samples (this work, Li22, and Li23) reveal remarkably consistent trends, supporting the robustness of our selection procedure.

\section{Modeling possible origins of the Sgr high-velocity star candidates} \label{sec:origin}

\subsection{The Hills mechanism} \label{subsec:Hills}
In this subsection, we investigate the hypothesis that HVSs can be produced at the center of the Sgr dSph via the Hills mechanism. We employed the Python package \textit{speedystar}\footnote{\url{https://github.com/fraserevans/speedystar}} \citep{Evans2022b} to generate HVS samples from the Hills mechanism, releasing them at the corresponding orbital positions of the Sgr dSph, integrating their orbits within the Galactic potential and performing mock observation. Our study focuses on exploring a parameter space defined by three key variables: the assumed mass of the central black hole in the Sgr dSph ($M_{BH}$), the initial mass function (IMF) slope ($\kappa$), and the HVS ejection rate ($\eta$).

It is well established that the stellar velocity dispersion ($\sigma$) and the mass of a central black hole ($M_{BH}$) are strongly correlated \citep{Kormendy2013, Greene2020}. Although the Sgr dSph is undergoing violent tidal disruption and lacks a classical bulge, its remnant velocity dispersion remains remarkably constant at $\sigma \approx 12\text{--}14~\mathrm{km~s^{-1}}$ \citep{Vasiliev2020}. To estimate the black hole mass, we considered two versions of the $M_{BH}$--$\sigma$ relation: the canonical relation inferred from a large sample of galaxies \citep{Kormendy2013}, where $M_{BH} \propto \sigma^{4.4}$, and a version specifically tailored to low-mass black holes \citep{Xiao2011}, where $M_{BH} \propto \sigma^{3.3}$. These relations yield a putative IMBH mass for the Sgr dSph within the range of $10^3\text{--}10^4~M_{\odot}$. Furthermore, as shown by the analysis of the NewHorizon simulation \citep{Dubois2021} in \cite{Beckmann2023}, dwarf galaxies with stellar masses similar to that of the Sgr dSph are expected to possess IMBHs that remain in their seed state ($M_{BH} < 10^4~M_{\odot}$). Consequently, we varied the parameter $M_{\mathrm{BH}}$ using values of $[10^3, 5 \times 10^3, 10^4]~M_{\odot}$ to encompass the physically plausible range for such a system.

There have been several independent lines of evidence suggesting that the IMF in the Sgr dSph is top-light. From chemical abundance analysis of stars in the Sgr dSph, \citet{McWilliam2013} identified a $\sim$0.4 dex deficiency in hydrostatic $\alpha$-elements (O and Mg) relative to explosive $\alpha$-elements (Si, Ca, and Ti), concluding that Sgr lacked enrichment from stars with masses $\gtrsim 30\,M_{\odot}$. Furthermore, chemical evolution modeling of Sgr stream stars by \citet{Carlin2018} suggests that a steeper power-law index ($\kappa \approx 2.75$) is necessary to match observed abundance patterns. Consequently, to encompass both the standard Salpeter slope and the steeper values suggested by these empirical constraints, we adopted a power-law form $f(m)\propto m^{-\kappa}$ with $\kappa = [2.35, 2.55, 2.75, 2.95, 3.15, 3.35]$.

The HVS ejection rate, $\eta$, is a relatively unconstrained parameter in our model. Theoretical estimates of the classical Hills mechanism in the GC typically range from $10^{-5}$ to $10^{-3}~\mathrm{yr}^{-1}$ \citep{Hills1988, Yu2003}, depending on the rate at which stars are scattered into the ``loss cone''---the region of phase space characterized by low-angular momentum orbits around $\mathrm{Sgr~A}^*$. Subsequent detailed simulations \citep[e.g.,][]{Zhang2013}, as well as calibrations based on the known population of HVS candidates \citep{Bromeley2012, Brown2014, Marchetti2018} and the frequency of tidal disruption events in the local Universe \citep{Bromeley2012, Brown2015}, yield consistent results. However, unlike the Milky Way center, the Sgr dSph is a system far from dynamical equilibrium due to ongoing tidal disruption. Since a self-consistent determination of the ejection rate through detailed dynamical simulations is beyond the scope of this work, we explored a wide range of values, $\eta \in [10^{-2}, 10^{-3}, 10^{-4}, 10^{-5}, 10^{-6}]~\mathrm{yr}^{-1}$, to encompass the plausible extremes and account for the significant uncertainties inherent to this disrupted system.

Other binary properties—such as the mass-ratio distribution $f(q)\propto q^\gamma$ and the semimajor axis distribution $f(a)\propto a^\alpha$—were shown by \cite{Evans2022a} to have negligible impact on the number of observable HVSs, and our experiments confirm insensitivity to $\gamma$ and $\alpha$ in the Sgr dSph Hills scenario. Therefore, the default value $\gamma=0$ and $\alpha=-1$ are taken by recommendation from \textit{speedystar}. We also must specify a metallicity distribution for the ejected stars—though our tests indicate negligible sensitivity of the detectable sample size to metallicity. Since there is currently no strong observational or theoretical evidence \citep{Wrobel2011, Beckmann2023} indicating that the putative black hole at the center of the Sgr dSph must reside precisely at its dynamical center and coincide with the globular cluster M54, we adopted the metallicity distribution of the Sgr dSph core from \cite{Mucciarelli2017} for our simulation. This distribution was modeled as a normal distribution with $\langle[\mathrm{Fe/H}]\rangle = -0.52$ dex and $\sigma = 0.17$ dex.

Using a Monte Carlo approach with the above parameters, we randomly generated HVSs and assigned each a flight time uniformly distributed between 0 and 100 Myr. We excluded stars with flight time exceeding 100 Myr, as they travel beyond observable distances. We further accounted for prior main-sequence evolution by assuming each HVS has already consumed a portion of its main-sequence lifetime before ejection. Any star whose remaining main-sequence lifetime is shorter than its flight time was excluded from the simulation (we limited our analysis to unevolved stars and did not consider the evolved population for detection).

The above procedures yielded an HVS sample with known present‐day positions, velocities, radii, luminosities, and effective temperatures. We computed their absolute magnitudes in the Gaia photometric bands and astrometric uncertainties following the method of \cite{Evans2022b}, based on each star’s effective temperature, luminosity, and surface gravity, and accounting for line‐of‐sight extinction using the Galactic dust map \texttt{combined15} of \textit{mwdust}
\footnote{\url{https://github.com/jobovy/mwdust}}
\citep{Bovy2016}. We then estimated apparent magnitudes in the Gaia and Johnson–Cousins systems via the MIST model grids \citep{Choi16, Dotter2016}, and derived $G_\mathrm{RVS}$ magnitudes from polynomial fits of \cite{Jordi2010} using Gaia $G$ and Johnson–Cousins $V$ and $I_c$ bands.

\begin{figure}[ht!]
    \centering
    \includegraphics[width=\linewidth]{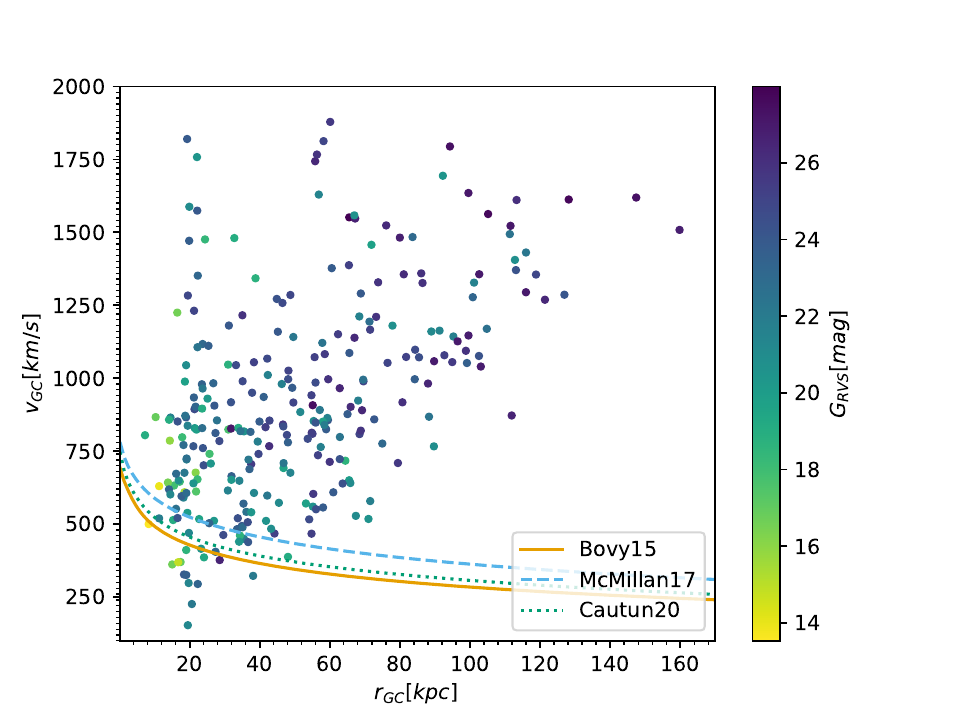}
    \caption{Distribution of galactocentric total velocity ($v_{GC}$) versus galactocentric radius ($r_{GC}$) for HVSs ejected via the Hills mechanism from the center of the Sgr dSph, color coded by each star's apparent magnitude in the Gaia RVS band. Results generated using the \textit{speedystar} code assuming $M_{bh}=10^4 M_\odot$, ejection rate $\eta=10^{-4} yr^{-1}$ and a Salpeter IMF \citep{Salpeter1955} with $\kappa=2.35$. Solid, dashed, and dotted curves represent the escape-velocity profiles predicted by three Milky Way potential models (\citealt{Bovy2015}; \citealt{McMillan2017}; \citealt{Cautun2020}).}
    \label{fig:Sgr_Hills}
\end{figure}

In Figure \ref{fig:Sgr_Hills}, we show the distribution of total velocities of Hills mechanism ejected HVS from a putative black hole at the center of the Sgr dSph, plotted in the Galactic‐center rest frame as a function of galactocentric distance. In this simulation, we adopt a central black hole mass of $10^4 M_\odot$, an HVS ejection rate of $10^{-4} \mathrm{yr}^{-1}$, and an initial mass function given by the classical Salpeter law \citep{Salpeter1955} with $\kappa=2.35$. The resulting HVS span a very broad radial range—from the Solar neighborhood out to beyond 100 kpc—with most stars exceeding the local escape‐velocity curve. This implies that, if HVSs are produced by the Hills mechanism in the Sgr dSph, their velocities would likely lie well above the escape speed, reaching on the order of $10^3 \mathrm{km~s^{-1}}$. We also color‐code each HVS by its $G_\mathrm{RVS}$ magnitude, since this band places the most stringent observational limits: Gaia DR3 RVS sample requires $G_\mathrm{RVS}<14$, whereas Gaia DR4 RVS sample will extend to $G_\mathrm{RVS}<16.2$. Under this model, most simulated HVSs are too faint in the RVS band to be detected by Gaia DR3, and only a small portion would be accessible to Gaia DR4.

In Gaia DR3 and DR4, an HVS must satisfy $G<20.7$ to be detectable. Since Gaia's RVS sample remains the most efficient channel for identifying potential HVSs from the Sgr dSph (60 out of 95 candidates in this work are from the Gaia DR3 RVS sample), it is important to estimate how many HVSs can be included in the Gaia RVS sample. In the following, when we refer to an HVS as “observable” in Gaia DR3/DR4, we mean that it can be included in the corresponding RVS sample. Membership in the DR3 RVS sample further requires $G_{\mathrm{RVS}}<14$ for stars cooler than 6900 K or $G_{\mathrm{RVS}}<12$ for stars cooler than 14,500 K. In DR4, the requirements are relaxed to $G_{\mathrm{RVS}}<16.2$ for stars cooler than 6900 K or $G_{\mathrm{RVS}}<14$ for stars cooler than 14,500 K.

To estimate astrometric errors, we adopted the five-dimensional covariance distribution function for DR3 measurements proposed by \citet{Everall2021}, and projected anticipated DR4 improvements by scaling down the DR3 errors. Radial-velocity uncertainties were estimated using \textit{PyGaia}\footnote{\url{https://github.com/agabrown/PyGaia}}
, based on each star's effective temperature and $G_{\mathrm{RVS}}$. To distinguish HVSs from field stars, we further required each star's total velocity to exceed the local escape velocity. Finally, to ensure precise orbit determinations, we demanded that the parallax uncertainty be less than 20\% of the measured parallax.

For each parameter combination, we performed 100 simulations and applied 1000 bootstrap resamplings. For each bootstrap resample, we computed the mean number of observed HVSs, $N_\mathrm{HVS}$. The median (50th percentile) of these resampled means was adopted as the central value, with uncertainties given by the 16th and 84th percentiles.

\begin{figure*}[ht!]
    \centering
    \includegraphics[width=\linewidth]{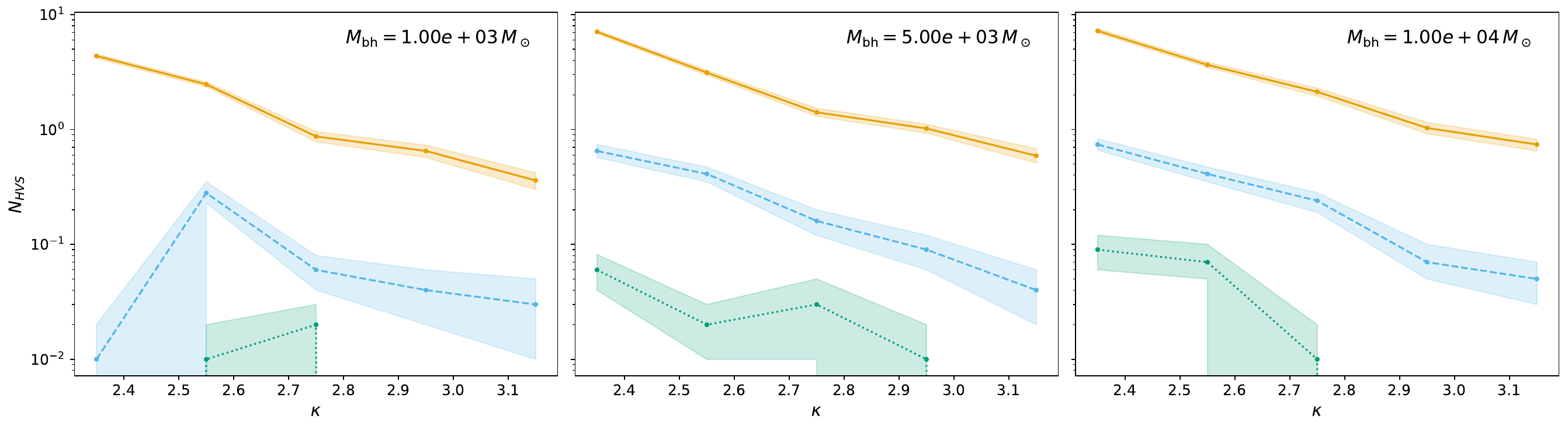}
    \caption{Predicted number of observable HVSs, $N_\mathrm{HVS}$, in Gaia DR4 as a function of the IMF slope $\kappa$. Panels show results for different assumed central black hole masses in the Sgr dSph: $M_{bh}=10^3M_\odot$ (left), $5\times10^3M_\odot$ (center), and $10^4M_\odot$ (right). Solid orange, dashed sky-blue, and dotted bluish-green lines correspond to ejection rates of $\eta=10^{-2},10^{-3},\text{and}~10^{-4} \mathrm{yr}^{-1}$, respectively. Shaded regions indicate the 16th–84th percentile range from bootstrap resampling.}
    \label{fig:Hills_NHVS}
\end{figure*}

We find that $N_{\mathrm{HVS}}$ remains nearly zero across all parameter sets when considering Gaia DR3 observational constraints, as well as for cases with $\eta \leq 10^{-5}~\mathrm{yr}^{-1}$ in the Gaia DR4 mock catalog. Consequently, Figure \ref{fig:Hills_NHVS} only presents the results for the Gaia DR4 sample with $\eta \geq 10^{-4}~\mathrm{yr}^{-1}$, where the predicted number of HVSs is statistically significant. 

In DR4, $N_\mathrm{HVS}$ decreases significantly with increasing IMF slope $\kappa$. This trend reflects the transition from a canonical Salpeter to a top-light IMF, which reduces the number of high-mass stars bright enough to be detected in DR4. We also find a modest increase in $N_\mathrm{HVS}$ with black hole mass, likely because a more massive black hole imparts higher ejection velocities, allowing HVSs to reach the solar neighborhood. Moreover, changing $\eta$ by an order of magnitude shifts $N_\mathrm{HVS}$ by a similar factor. Our results agree with those of \citet{Evans2022a}: only when $\eta$ is high and the IMF is not top-light can a detectable sample be obtained. By adopting the proposed IMF slope of $\kappa = 2.75 \pm 0.2$ \citep{Carlin2018}, we suggest that it is unlikely to detect genuine HVSs in Gaia DR4 produced via the Hills mechanism, except under conditions of extremely high ejection rates ($\eta \geq 10^{-2} \mathrm{yr}^{-1}$).

\subsection{Tidal disruption of the Sgr dSph} \label{subsec:Tidal}

In this subsection, we discuss whether the tidal disruption of the Sgr dSph could be a possible channel to produce observed high-velocity stars that have a close encounter with the Sgr dSph. In the pioneering work, \citet{Abadi2009} first proposed that stars stripped from a disrupted satellite galaxy could achieve velocities comparable to those produced by the Hills mechanism. Building upon this idea, \citet{Piffl2011} studied the scenario in detail by performing a suite of $N$-body simulations of a satellite’s encounter with a Milky Way–type host. They gathered statistics on the properties of stripped stars and showed that the maximum energy kick imparted to tidal debris depends primarily on the satellite’s initial mass. Their results indicate that only massive satellites ($M_{\rm sat}>10^9\,M_\odot$) can produce unbound stellar populations via tidal stripping.

To predict the kinematics of stripped stars, \citet{Piffl2011} developed a simple analytical model to estimate the maximum energy imparted to stars during the tidal disruption of the satellite. Applying this model to the Sgr dSph, which has a pericentric speed $V_{\rm peri}\approx350\ \mathrm{km\,s^{-1}}$, an escape speed $v_{\rm esc}(R_{\rm tidal})\approx40\ \mathrm{km\,s^{-1}}$, and an orbital angular momentum $L_{\rm sat}\approx5400\ \mathrm{kpc\,km\,s^{-1}}$ \citep{Vasiliev2021}, one obtains $\Delta E_{\rm max}\approx9000\ \mathrm{km}^2\,\mathrm{s^{-2}}$. This value is much smaller than the magnitude of Sgr dSph’s orbital energy $E_{\rm sat}\approx-4.9\times10^4\ \mathrm{km}^2\,\mathrm{s^{-2}}$. To further validate the results presented above, we have constructed an analytical model that captures the detailed physical processes governing stellar stripping from the Sgr dSph. In this model, stars initially bound to Sgr dSph orbit its center with speed $\mathbf{v}_{\rm ini}$. As the satellite galaxy plunges through the Milky Way’s potential, its stars experience impulsive energy injections from the Galactic tidal field. Once a star acquires sufficient energy, it will escape from the Sgr dSph with a final speed of
\begin{equation}\label{eq:impulse_escape}
    \mathbf{v}_{\rm fin} = \mathbf{v}_{\rm ini} + \Delta \mathbf{v},
\end{equation}
where $\Delta \mathbf{v}$ represents the velocity impulse imparted by the tidal interaction. For a fixed magnitude of $\Delta \mathbf{v}$, the magnitude of the final velocity $|\mathbf{v}_{\rm fin}|$ is maximized if and only if $\Delta \mathbf{v}$ is parallel to $\mathbf{v}_{\rm ini}$. To quantify $\Delta \mathbf{v}$, we computed the energy input using the so-called “tidal shock” formalism. In this framework, the change in a star’s kinetic energy is determined by integrating the time-varying tidal acceleration it experienced along the Sgr dSph’s orbit. We discuss the tidal shock approximation below.

The tidal shock formalism rests on two fundamental hypotheses. The first, known as the “impulse approximation,” posits that the internal velocity dispersion of the subject system is negligible compared to its relative velocity at pericenter passage. In the case of the Sgr dSph, the velocity dispersion, $\sigma_v \approx 13\ \mathrm{km\ s^{-1}}$ \citep{Vasiliev2020}, is indeed much smaller than its pericentric speed around the Milky Way, $V_{\rm peri} \approx 350\ \mathrm{km\ s^{-1}}$. Under this approximation, during pericenter passage (when tidal force is strongest) the stars within Sgr dSph can be treated as effectively stationary with respect to the galaxy’s center, allowing the tidal force to act impulsively.

The second fundamental hypothesis, known as the ``distant-tide approximation", states that when the separation between the perturber (the Milky Way, in this case) and the subject system (the Sgr dSph) greatly exceeds the subject system’s characteristic size, the external potential varies smoothly across the system. Under this condition, one may expand the perturber’s gravitational field about the subject system’s center in a Taylor series. In a frame where the Sgr dSph's center of mass is at the origin, let $\Phi(\mathbf{x}, t)$ denote the gravitational potential of the MW. The tidal acceleration of star $\alpha$ relative to the origin is given by: 
\begin{equation}
    \label{eq:distant_tide_acceleration}
    \dot{\mathbf{v}}_\alpha = -\sum_{j,k=1}^3\hat{e}_j\Phi_{jk}x_{\alpha k},
\end{equation}
where \begin{equation} \label{eq:_distant_tide_2}
    \Phi_{jk}\equiv\frac{\partial^2\Phi}{\partial x_j\partial x_k}\bigg|_{\mathbf{x}=0}
\end{equation}
denotes the tidal tensor of the gravitational field induced by the perturber, evaluated at the origin of the Sgr-centric coordinate frame.

When both the distant-tide and high-speed (impulse) approximations hold simultaneously, the encounter is often referred to as a “tidal shock.” In the case of the Sgr dSph, the pericentric distance to the Milky Way is $r_{\rm peri}\approx16\ \mathrm{kpc}$, while its tidal radius is $r_{t}\approx2\ \mathrm{kpc}$ \citep{Vasiliev2020}. Although the strict inequality $r_{t}\ll r_{\rm peri}$ does not hold, we aim only to estimate the order of energy input from the tidal shock. Our estimated energy injection agrees well with the results of the $N$-body simulations presented by \citet{Vasiliev2021}, lending support to the validity of this simplified approach. The physical meaning of Equation (\ref{eq:distant_tide_acceleration}) is that the tidal acceleration on a star is correlated with the distance between it and the center of Sgr dSph, so we focus on stars at the tidal radius to evaluate the high-speed star generation by tidal disruption. For consistency with the preceding analysis, we adopted the \texttt{MWPotential2014} model \citep{Bovy2015} to compute the tidal forces acting on the Sgr dSph. This potential is composed of a spherical bulge characterized by a power-law density profile with an exponential cut-off, an axisymmetric Miyamoto–Nagai disk, and a spherical Navarro–Frenk–White halo.

Suppose the trajectory of Milky Way in the Sgr-centric frame is given by the closed expression  
$\mathbf{X}(t)$. At any instant, one can perform a straightforward exercise to compute the tidal tensor at the center of mass of the Sgr dSph. The resulting tidal acceleration of a star $\alpha$ depends on time only through the trajectory $\mathbf{X}(t)$. We first computed the orbit of the Sgr dSph over the time interval [-0.5, 0.5] Gyr—encompassing its most recent pericentric passage around the Milky Way using \textit{galpy} \citep{Bovy2015}.
Next, we performed spline interpolation on the orbit of Sgr dSph. We then numerically integrated the tidal acceleration of stars relative to the Sgr dSph centroid along this interpolated orbit to obtain the velocity increment $\Delta \mathbf{v}$. This computation was performed using the \textit{scipy} library \citep{Virtanen2020}.

\begin{figure*}[htbp]
    \centering
    \includegraphics[width=\linewidth]{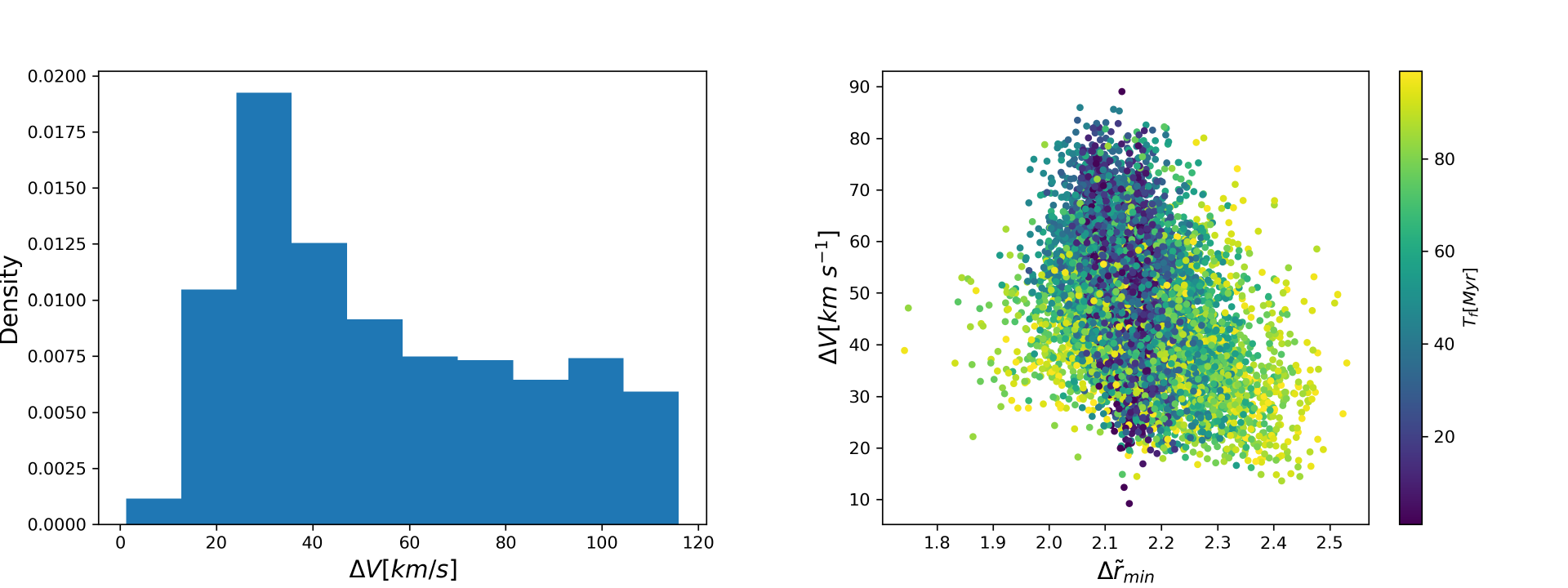}
    \caption{Left panel: Distribution of the impulse imparted to stars uniformly distributed on the tidal radius of the Sgr dSph. Right panel: Velocity difference between each star and the Sgr dSph’s center of mass at the stripping time, plotted against the normalized radial separation ($\Delta \Tilde{r}=r_{min}/R_h$), color coded by flight time.}
    \label{fig:tidal_dv}
\end{figure*}

We randomly placed 1000 points at the tidal radius of the Sgr dSph and then integrated their velocity increments $\Delta \mathbf{v}$ relative to the Sgr dSph’s center induced by the tidal forces.  The results are shown in the left panel of Figure \ref{fig:tidal_dv}.  We find that for most of the points at the tidal radius the velocity increment is approximately several tens $\mathrm{km~s^{-1}}$.  According to Equation (\ref{eq:impulse_escape}), if the direction of this increment aligns with the star’s initial velocity relative to the center of mass (typically $10-20~\mathrm{km~s^{-1}}$; see \citealt{Vasiliev2020}), stars attain the maximum escape velocity relative to the center, which is also below $150~\mathrm{km~s^{-1}}$.  This is far smaller than the $400-600~\mathrm{km~s^{-1}}$ relative velocities observed for the high-speed stars that have experienced a close encounter with the Sgr dSph.

These results can also be validated against the N-body model of \citet{Vasiliev2021}, which provides a snapshot of the Sgr dSph following its tidal disruption by the Milky Way.  We selected particles that were stripped within the last 100 Myr in this model and integrated their orbits backward to the moment they were tidally stripped.  The outcome is presented in the right panel of Figure \ref{fig:tidal_dv}. We find that most of these particles approach within 2.5 half-mass radii of the Sgr dSph, justifying our use of this radius as the criterion for identifying stars that have undergone a close encounter. Moreover, all of these particles reach maximum velocities below $100~\mathrm{km~s^{-1}}$ at pericenter, in good agreement with our estimates using the tidal-shock approximation. Therefore, for the Sgr dSph, tidal stripping alone is very unlikely to produce the high-velocity stars observed in our sample.

\subsection{Chance encounters of Milky Way halo stars with the Sgr dSph} \label{subsec:halo}

In this subsection, we investigated the scenario in which high-velocity stars are simply Milky Way halo stars that naturally pass by the Sgr dSph at its pericenter before moving into the solar neighborhood. To assess this possibility, it is essential to construct a robust model of Milky Way background stars. For this purpose, we adopted the model developed by \cite{Binney2023} (hereafter BV23 model).

The adopted framework is defined by parameters governing the action-based distribution functions $f(\mathbf{J})$ for four stellar disks (three thin-disk age cohorts and one thick disk), as well as for a spheroidal bulge and spheroidal components of both the stellar and dark halos. Given these distribution functions and a specified gas distribution, one can self-consistently solve for the spatial densities of stars and dark matter and the gravitational potential they generate. The first constraint of the BV23 model is the kinematics of stars from Gaia DR2 \citep{Gaia2018} that possess the Radial Velocity Spectrometer measured line-of-sight velocities \citep{katz2018}. The stellar locations and velocities of RVS stars used in the BV23 model are computed by \cite{Schonrich2019}. The second constraint is the vertical stellar density profile in the column above the Sun \citep{Gilmore1983}. The model is implemented using the \textit{Agama} code \citep{Vasiliev2019}.

We first sampled 100 million stars from the BV23 model, then selected stars with heliocentric distances less than 5 kpc, as parallax measurements are more reliable within this range. Next, we applied the same velocity cut as in Section \ref{sec:sample}, identifying stars whose galactocentric total velocity exceeds 400 $\mathrm{km~s^{-1}}$, which left us with 2408 stars. We then selected the subset of these stars with $r_{\mathrm{min}} < 2.5R_h$, leaving 81 stars in total. The fraction of these stars whose orbits trace back to the Sgr dSph relative to the total high-velocity sample is approximately 3.36\%, which closely matches the observed fraction (2.02\% for Gaia DR3 RVS, 2.32\% for DESI DR1, and 3.44\% for LAMOST DR12).

\begin{figure}[ht!]
    \centering
    \includegraphics[width=\linewidth]{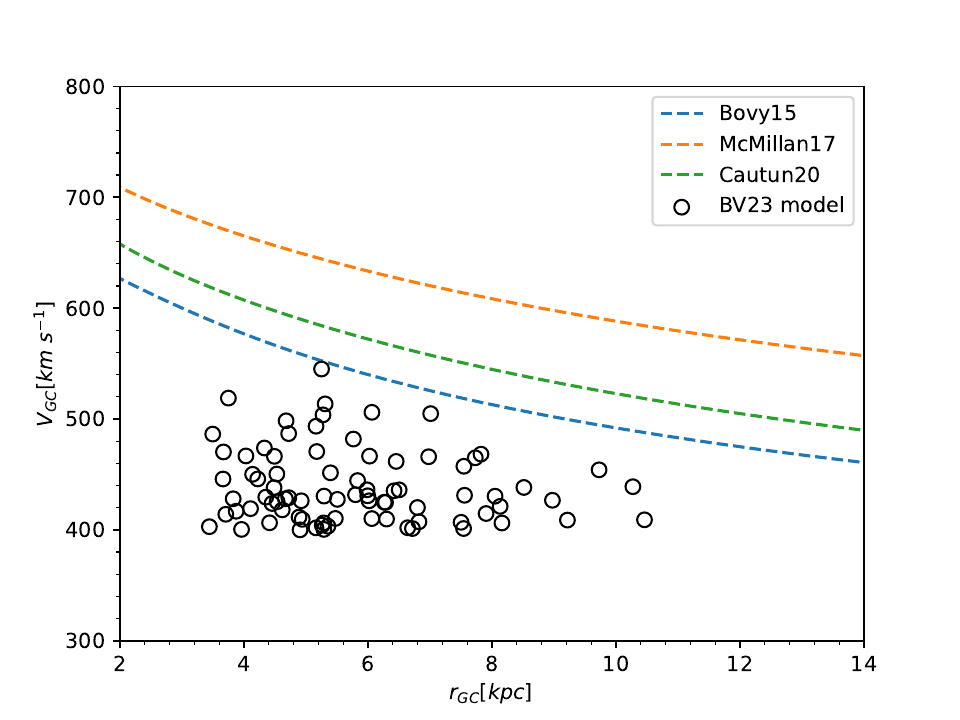}
    \caption{Distribution of galactocentric total velocity ($v_{\mathrm{GC}}$) versus galactocentric radius ($r_{\mathrm{GC}}$) for stars sampled from \cite{Binney2023} model with $v_{\mathrm{GC}} > 400 \mathrm{km~s^{-1}}$ and $r_{\mathrm{min}} < 2.5 R_h$. Solid, dashed, and dotted curves show the escape-velocity profiles predicted by three Milky Way potential models: \cite{Bovy2015}, \cite{McMillan2017}, and \cite{Cautun2020}.}
    \label{fig:scm_vgc_over_rgc}
\end{figure}

Figure \ref{fig:scm_vgc_over_rgc} presents the galactocentric total velocity ($v_{\mathrm{GC}}$) as a function of galactocentric radius ($r_{\mathrm{GC}}$) for the simulated stars that satisfy the aforementioned selection criteria. The simulated sample traces the same locus as the bulk of the observed stars in Figure \ref{fig:vgc_over_rgc}. Notably, none of the simulated stars exceeds the local escape speed predicted by any of the Galactic potentials adopted in this study \citep{Bovy2015, McMillan2017, Cautun2020}, which is because actions can only be computed for bound stars.

\begin{figure*}[ht!]
    \centering
    \includegraphics[width=\linewidth]{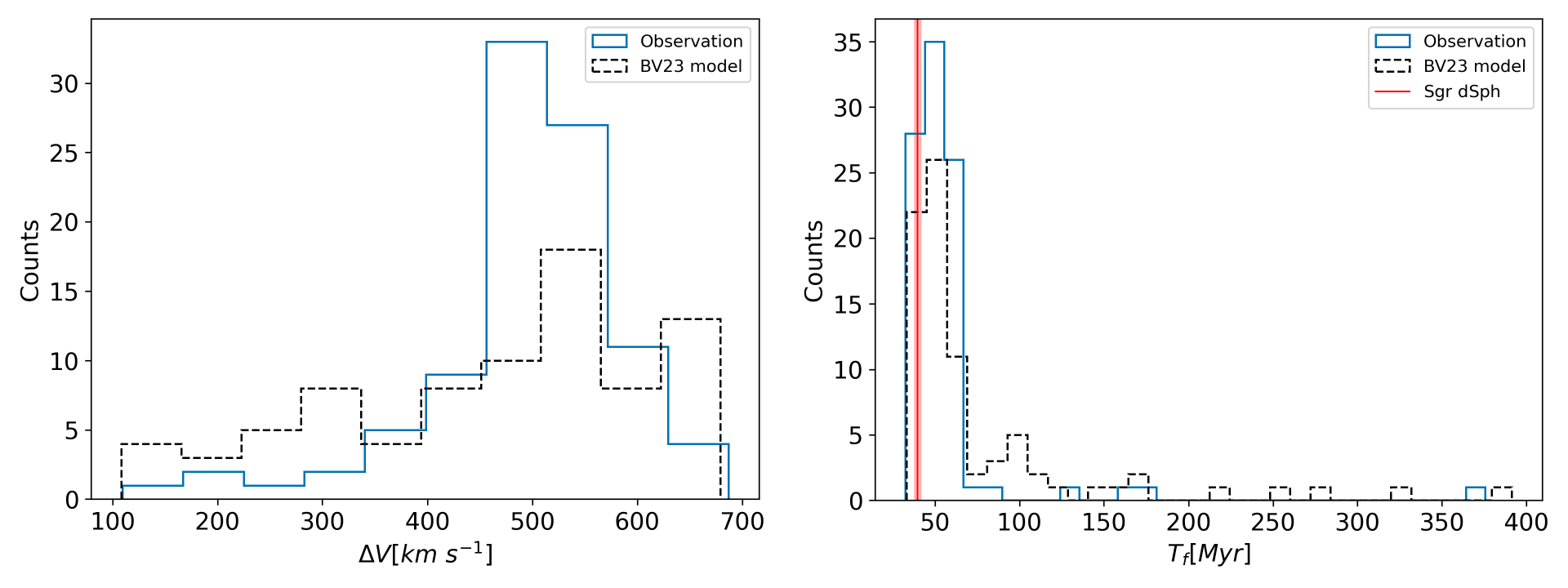}
    \caption{Left panel: Distribution of relative velocities at pericentric passage ($\Delta V$) for high-velocity stars encountering the Sgr dSph, with observations shown as solid blue lines and the BV23 model as dashed black lines.
    Right panel: Distribution of flight times since pericenter ($T_f$). The line styles are the same as in the left panel.}
        \label{fig:scm_dv_and_tf}
    \end{figure*}

In Figure \ref{fig:scm_dv_and_tf}, we present the relative velocities at pericentric passage and the corresponding flight times for high-velocity stars in the BV23 model that undergo close encounters with the Sgr dSph. The left panel shows the distribution of relative velocities, while the right panel shows the distribution of flight times, and for comparison we overplot the observed distributions of high-velocity stars that encounter the Sgr dSph. The ranges of both relative velocity and flight time in the BV23 model closely match the observations, and the model also reproduces the “low-velocity tail” and the “long-flight-time tail.” This agreement indicates that the model captures the underlying physical processes in the data rather than merely reflecting selection effects. 

\begin{figure*}[ht!]
    \centering
    \includegraphics[width=\linewidth]{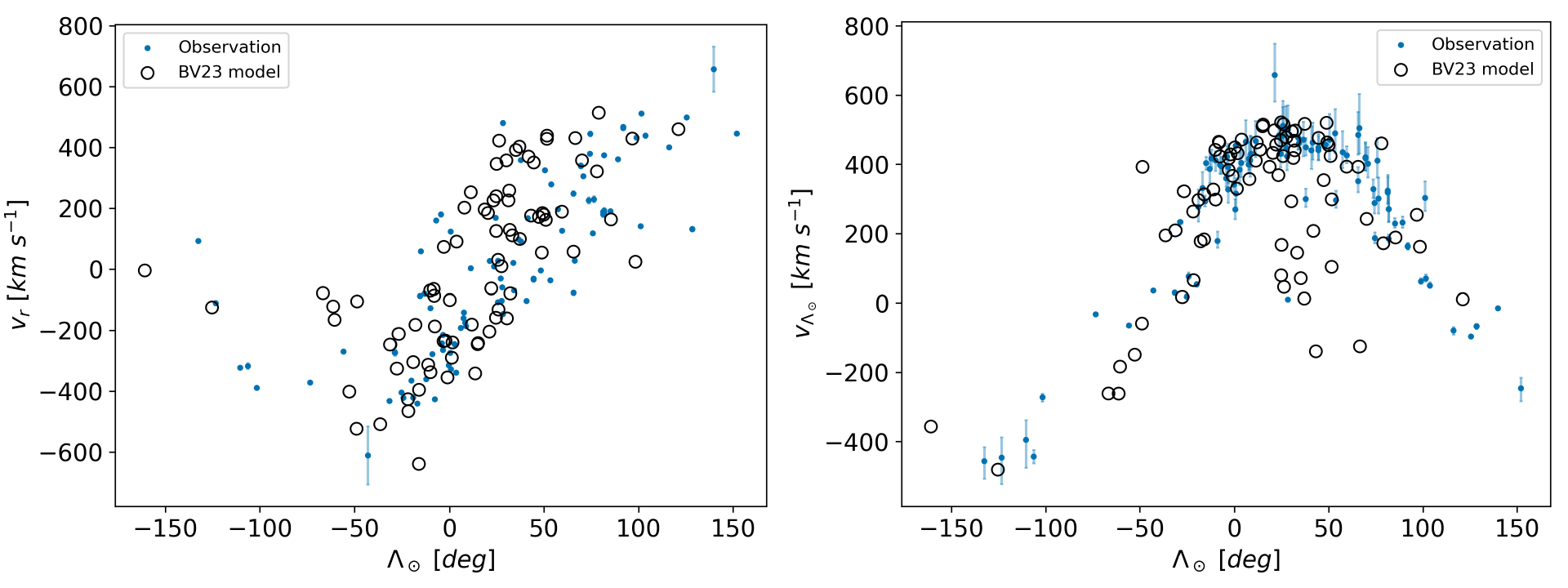}
    \caption{Distribution of high-velocity stars in the Sgr-stream coordinate system. Left panel: Radial velocity $v_{r}$ as a function of $\Lambda_\odot$. Right panel: Tangential velocity along the stream longitude, $v_{\Lambda_\odot}$, versus $\Lambda_\odot$. The black circles represent stars from the BV23 model, and blue dots indicate the observed high-velocity stars that encounter the Sgr dSph in this work.}
    \label{fig:scm_rv_and_vL_over_L}
\end{figure*}

In Figure \ref{fig:scm_rv_and_vL_over_L}, we show the distribution of high-velocity stars in the BV23 model that undergo close pericentric encounters with the Sgr dSph, expressed in the Sgr-stream coordinate system. The left panel plots radial velocity $v_{r}$ as a function of $\Lambda_\odot$, while the right panel shows the tangential velocity along the stream longitude, $v_{\Lambda_\odot}$, versus $\Lambda_\odot$. The BV23 model captures the main kinematic features seen in the observations, including the positive correlation between $v_{r}$ and $\Lambda_\odot$ and the cosine-like variation of $v_{\Lambda_\odot}$ with $\Lambda_\odot$.  

To further quantify the statistical consistency between the observations and the BV23 model, we performed a series of nonparametric tests. Two-sample Kolmogorov–Smirnov (KS) tests yield $p$-values of 0.37 for the relative velocities and 0.36 for the flight times. Furthermore, a two-sample energy test \citep{Sz2013Energy} yields $p$-values of 0.33 for the $v_r$--$\Lambda_\odot$ distribution and 0.10 for the $v_{\Lambda_\odot}$--$\Lambda_\odot$ projection. These high $p$-values indicate that the observed and modeled distributions are statistically indistinguishable.

Combining the discussion in this subsection with the previous two, we find that for the Sgr dSph, the two mechanisms traditionally considered as potential channels for producing high-velocity stars—the Hills mechanism and tidal disruption—both face difficulties in explaining the observed stars whose orbits trace back to the Sgr dSph. High-velocity stars produced by the Hills mechanism are unlikely to be detected with current observational facilities, while tidal disruption is theoretically inefficient in generating such stars. By contrast, our analysis of the BV23 model shows that chance encounters of Milky Way halo stars with the Sgr dSph yield a distribution that closely resembles the observations. This suggests that the observed high-velocity stars are most likely halo stars of the Milky Way.
Furthermore, this provides a natural explanation for why adopting a smaller distance to the Sgr center results in a higher count of candidate stars, as noted in Section \ref{subsec:orbits}. A smaller assumed distance implies that the Sgr dSph penetrates deeper into the Milky Way (reaching smaller galactocentric radii) during its pericentric passage. This deeper penetration into the denser regions of the Galactic halo facilitates a larger number of chance encounters with halo stars, thereby increasing the number of identified candidates.

It must be emphasized, however, that our discussion here is preliminary and strongly model-dependent. First, the BV23 model is an $N$-body simulation, and we have not performed mock observations to account for uncertainties in the phase-space coordinates of stars. Such a procedure requires consideration of the stellar mass distribution, metallicity, dust extinction, and telescope selection functions across the Galaxy, which is a highly nontrivial task that we plan to address in future work. Second, while the BV23 model provides a relatively robust description of the Galactic disk, it may be insufficient for the halo, where the influence of the Gaia-Sausage-Enceladus merger likely needs to be considered. Using other Galactic models could lead to different conclusions. For example, when we employed the widely used GEDR3 mock model \citep{Rybizki2020}, we found only about ten stars in the entire dataset whose orbits intersect the Sgr dSph. A detailed comparison and discussion of different Galactic models will be presented in future work.

\section{Chemical analysis}\label{sec:Chemical}

In this section, we conducted a complementary chemical abundance analysis to further substantiate our findings. Among our 95 high-velocity star candidates, 17 from DESI DR1 and 18 from LAMOST DR12 have metallicity measurements. However, the [$\alpha$/Fe] measurements from DESI exhibit data quality issues, as crossmatching with APOGEE DR17 reveals significant scatter and systematic biases. Furthermore, only a limited subset of the LAMOST sample has [$\alpha$/Fe] measurements. Consequently, we restricted our chemical analysis to [Fe/H].

\begin{figure}[htbp]
    \centering
    \includegraphics[width=\linewidth]{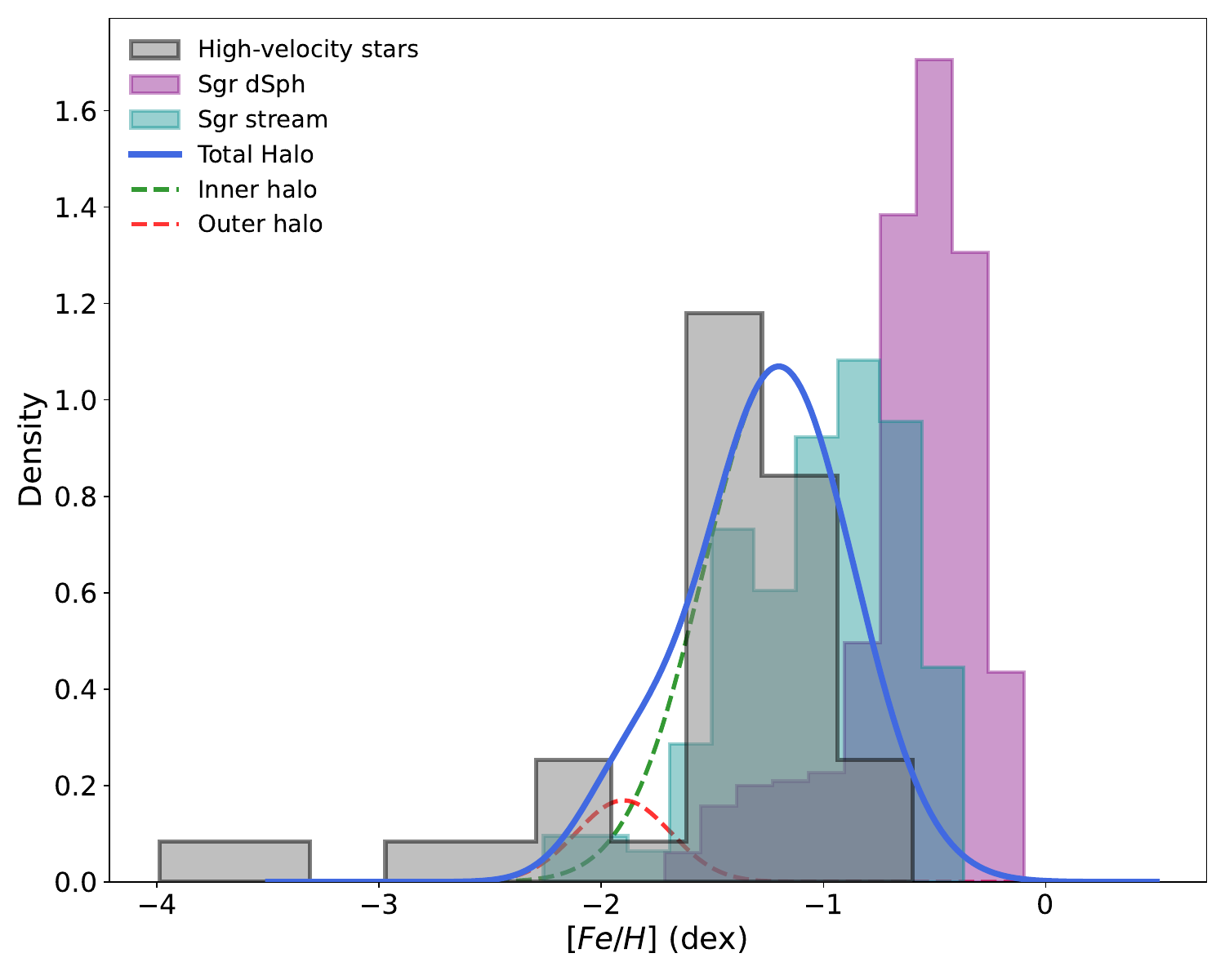}
    \caption{Metallicity ([Fe/H]) distribution of the 95 high-velocity star candidates compared with Sgr and Milky Way halo populations. The gray histogram shows our candidates. Sgr stream members (dark cyan shading) and Sgr dSph core members (purple shading) are from \cite{Hayes2020}. Milky Way halo giant stars from \cite{Liu2018} are shown with a solid blue line, encompassing the inner (dashed green) and outer (dashed red) halo components.}
    \label{fig:chemo}
\end{figure}

Figure \ref{fig:chemo} displays the metallicity distribution of our high-velocity candidates, which exhibits a bimodal structure with peaks at [Fe/H] $\sim -1.3$ and $\sim -2.1$ dex. For comparison, we plot the metallicity distributions of the Sgr dSph and its stream from \cite{Hayes2020}, which include 166 stream members and 710 members of the Sgr dSph core. The Sgr stream members peak at [Fe/H] $\sim -1$, while the Sgr core members peak at $\sim -0.5$, reflecting the metallicity gradient of the progenitor prior to tidal disruption. We also compare our sample with the Milky Way halo population characterized by \cite{Liu2018} using 4,680 giant stars ($|z| > 5$ kpc) from LAMOST. Their study describes the halo metallicity profile as a bimodal Gaussian with peaks at [Fe/H] $\sim -1.2 \pm 0.3$ (inner halo) and $-2.0 \pm 0.2$ (outer halo), comprising 91\% and 9\% of the population, respectively.

The chemical composition provides a crucial test for the origin of these stars. Notably, the bulk of the Sgr stream was stripped during the most recent pericentric passage approximately 1 Gyr ago. Li22 and Li23 invoked the chemical similarity between high-velocity stars and the Sgr stream as evidence to support a Sgr origin. However, our high-velocity candidates experienced close encounters with the Sgr dSph within the last 100 Myr, a much shorter timescale than the typical stripping history of the stream. If these stars indeed originated from the Sgr dSph, they would have had to be ejected by a violent dynamical process within the core. Nevertheless, the metallicity of our high-velocity sample deviates significantly from that of the Sgr core.

Furthermore, a two-sample KS test yields a p-value of $5.38 \times 10^{-7}$ when comparing the Sgr stream members to our sample, strongly rejecting the hypothesis that these two populations are drawn from the same distribution. As shown in Figure \ref{fig:chemo}, while there is overlap in the metallicity range, the distribution peaks are distinct: the Sgr stream is metal-richer by $\sim 0.2$--$0.3$ dex compared to both our candidates and the Milky Way halo. Conversely, a two-sample KS test between our sample and the halo distribution yields a p-value of 0.019. Given that significance thresholds are statistical conventions without intrinsic physical meaning, the key point is that this p-value is orders of magnitude higher than the result for the Sgr stream (for comparison, the KS test between the Sgr stream and the Milky Way halo yields $p=7.35 \times 10^{-14}$).

However, we identify three very metal-poor (VMP) stars with [Fe/H] below $-2.5$ dex that cannot be adequately represented either by the Sgr or Milky Way halo distributions. These stars have metallicities of $-3.48 \pm 0.08$ dex, $-3.99 \pm 0.04$ dex, and $-2.71 \pm 0.09$ dex, warranting further discussion. As noted by \cite{Liu2018}, their halo sample is relatively small and incomplete at the very metal-poor end. After excluding these three stars from our high-velocity sample, the p-value between the truncated sample and the Milky Way halo increases to 0.056, while the p-value between the truncated sample and the Sgr stream increases to $2.77 \times 10^{-6}$.

All three VMP stars originate from DESI DR1, with G-band magnitudes of 19.88 mag, 19.15 mag, and 17.46 mag, respectively. Given that LAMOST observations are challenging at such faint magnitudes, these stars are not well represented by the Milky Way metallicity distribution from \citet{Liu2018}, which is based on LAMOST data. In contrast, DESI provides an order-of-magnitude increase in the number of faint stars with $17.5 < G < 21$ that have radial velocity and abundance measurements compared to existing spectroscopic surveys, making it highly suitable for identifying faint VMP stars \citep{Koposov2025}. While a more comprehensive Milky Way metallicity distribution extending to the very-metal-poor end could be constructed from DESI data, this investigation is beyond the scope of the present work.

\section{Discussion and conclusions} \label{sec:D&C}

In this work, we used Gaia DR3, DESI DR1, and LAMOST DR12 to identify 95 high-velocity stars with close encounters to the Sgr dSph. Their relative velocities ($\Delta V \sim 400$–$600\ \mathrm{km~s^{-1}}$) and short flight times ($T_f < 100\ \mathrm{Myr}$) coincide with the epoch of Sgr’s recent pericentric passage, and their kinematics follow coherent trends in Sgr stream coordinates.

We systematically tested the three possible origins proposed in previous studies. Simulations show that the Hills mechanism would not yield detectable HVSs in Gaia DR3 and only marginally in DR4 under extreme assumptions, while tidal disruption provides velocity kicks ($\Delta V < 150\ \mathrm{km~s^{-1}}$) far below the observed range. By contrast, the \cite{Binney2023} halo model reproduces the observed velocity and orbital distributions, indicating that most candidates are ordinary halo stars. This conclusion is further supported by our metallicity distribution analysis, which indicates that these high-velocity stars are chemically consistent with the Milky Way halo population.

There are other mechanisms capable of producing high-velocity stars that were not explicitly discussed in previous sections. First, a supernova origin can be confidently excluded because the gas in the Sgr dSph was stripped approximately 1 Gyr ago \citep{Tepper2018}, consistent with the last substantial burst of star formation in Sgr. This time interval significantly exceeds the lifespan of massive stars that undergo core-collapse. Second, for the same reason, the Sgr dSph possesses no young massive clusters. Third, the Sgr dSph currently hosts four bound globular clusters: M54, Terzan 7, Terzan 8, and Arp 2, with M54 located coincident with the core of the Sgr dSph. While single star-binary interactions involving compact objects in globular clusters can eject high-velocity stars \citep{Weatherford2023, Evans2025}, the ejection rate for stars with velocities exceeding $400~\mathrm{km~s^{-1}}$ (as observed in this work) is two or three orders of magnitude lower than that of the Hills mechanism \citep{Huang2025, Evans2025}, and is therefore negligible.

Thus, our results suggest that both the candidates identified in this work and those reported in earlier studies through orbital back-tracing to the Sgr dSph are most likely ordinary halo stars. This highlights the need for caution when inferring the origin of high-velocity stars solely from orbital reconstruction, as the contribution of halo stars must be carefully accounted for. Future studies may benefit from incorporating more detailed chemical abundance information to obtain more robust conclusions. Since halo stars are, by definition, bound to the Galaxy, a secure detection of truly unbound HVSs originating from the Sgr dSph would immediately point to intriguing and extreme dynamical processes—though no such sources are found in this study. While our analysis indicates that genuine Sgr-origin HVSs are unlikely to be identified under current observational limits, this work provides a useful stepping stone for future searches.

\begin{acknowledgements}
This work was supported by National Key R\&D Program of China No. 2024YFA1611900, and the National Natural Science Foundation of China (NSFC Nos. 11973042, 11973052). 

This work has made use of data from the European Space Agency (ESA) mission Gaia (\url{https://www.cosmos.esa.int/gaia}), processed by the Gaia Data Processing and Analysis Consortium (DPAC, \url{https://www.cosmos.esa.int/web/gaia/dpac/consortium}). Funding for the DPAC has been provided by national institutions, in particular the institutions participating in the Gaia 
Multilateral Agreement. 

This research used data obtained with the Dark Energy Spectroscopic Instrument (DESI). DESI construction and operations is managed by the Lawrence Berkeley National Laboratory. This material is based upon work supported by the U.S. Department of Energy, Office of Science, Office of High-Energy Physics, under Contract No. DE–AC02–05CH11231, and by the National Energy Research Scientific Computing Center, a DOE Office of Science User Facility under the same contract. Additional support for DESI was provided by the U.S. National Science Foundation (NSF), Division of Astronomical Sciences under Contract No. AST-0950945 to the NSF’s National Optical-Infrared Astronomy Research Laboratory; the Science and Technology Facilities Council of the United Kingdom; the Gordon and Betty Moore Foundation; the Heising-Simons Foundation; the French Alternative Energies and Atomic Energy Commission (CEA); the National Council of Humanities, Science and Technology of Mexico (CONAHCYT); the Ministry of Science and Innovation of Spain (MICINN), and by the DESI Member Institutions: www.desi.lbl.gov/collaborating-institutions. The DESI collaboration is honored to be permitted to conduct scientific research on I’oligam Du’ag (Kitt Peak), a mountain with particular significance to the Tohono O’odham Nation. Any opinions, findings, and conclusions or recommendations expressed in this material are those of the author(s) and do not necessarily reflect the views of the U.S. National Science Foundation, the U.S. Department of Energy, or any of the listed funding agencies.

LAMOST is a National Major Scientific Project built by the Chinese Academy of Sciences, which has been provided by the National Development and Reform Commission.
\end{acknowledgements}

\bibliographystyle{aa}
\bibliography{myBib}
\end{document}